\documentclass[11pt, aps, amssymb,amsmath,showpacs,nofootinbib,floatfix]{revtex4}
\usepackage{epsfig}
\usepackage{graphicx}
\newcommand{\CM}{{\mathbb C}}
\newcommand{\Tr}{\mbox{\rm Tr}}
\newcommand{\NM}{{\mathbb N}}

\newcommand{\RM}{{\mathbb R}}
\newcommand{\SM}{{\mathbb S}}

\newcommand{\ZM}{{\mathbb Z}}

\newcommand{\real}{\Re\mbox{\rm e}}
\newcommand{\hosc}{H^{\mbox{\rm\tiny (osc)}}_{\omega}}

\newtheorem{proposi}{Proposition}
\newtheorem{lemma}{Lemma}
\newtheorem{coro}{Corollary}

\newcommand{\lgn}{\pmb\lVert}
\newcommand{\rgn}{\pmb\rVert}
\newcommand{\tp}{\ell^2(\NM)\otimes L^2_+(\SM)}
\newcommand{\exc}{{\cal E}_{\omega}^*}

\begin{document}

\title{Irreversible Behaviour and Collapse of Wave Packets in a Quantum System with Point Interactions.}
\author{Italo Guarneri
} \affiliation {
{\small  Center for Nonlinear and Complex Systems}\\
{\small  Universit\'a dell'Insubria, via Valleggio 11, I-22100 Como, Italy.}\\
{\small  Istituto Nazionale di Fisica Nucleare, Sezione di Pavia,
via Bassi 6, I-27100 Pavia, Italy.} }
\begin{abstract}{
A system of a particle and a harmonic oscillator, which have  pure point spectra if uncoupled, is known to acquire   absolutely continuous spectrum when they are coupled by a sufficiently strong point interaction. Here the  dynamical mechanism  underlying this spectral phenomenon is exposed.  The  energy of the oscillator is proven to exponentially diverge in time, while the spatial probability distribution of the  particle
collapses into a $\delta$-function in the interaction point. On account of this result, a generalized model with many oscillators which interact with the particle at different points is argued to provide a formal model for approximate measurement of position,
and collapse of wave packets.}
\end{abstract}
%\author{}
%\address
%{$^3$ Universit\`a dell'Insubria a Como, via Valleggio 11, 22100 Como,
%Italy \\

%$^4$ Istituto Nazionale per la Fisica della Materia,
%via Celoria 16, 20133 Milano, Italy\\
%$^5$ Istituto Nazionale di Fisica Nucleare, Sezione di Pavia,
%via Bassi 6, 27100 Pavia, Italy}

\date{\today}

%%%%%%%%% environments %%%%%%%%%%%%%%%%%%%%%%%%%%%%%%%
%\newtheorem{theo}{Theorem}
%\newtheorem{defini}{Definition}
%\newtheorem{proposi}{Proposition}
%\newtheorem{lemma}{Lemma}
%\newtheorem{coro}{Corollary}
%\newtheorem{rem}{Remark}
%\newtheorem{hypo}{Hypothesis}

%%%%%%%%% calligraphic %%%%%%%%%%%%%%%%%%%%%%%%%%%%%%%
%\twocol

%\draft: not for distribution.


\maketitle
%\centerline{\bf Preliminary draft - not for distribution.}
\section{Introduction.}
%\subsection{Quantum and classical irreversibility.}
\subsection{Background: Smilansky's  model.}
Smilansky's model \cite{smi04} is an offspring of the theory of Quantum Graphs. It consists of a quantum particle coupled to a harmonic oscillator via a point interaction. The particle moves inside  a 1-dimensional hard box . The linear coordinates of the harmonic oscillator and of the particle are  denoted by $q$ and by $x$ respectively , with $x\in I_L\equiv [-L/2,L/2]$. In the Hilbert space ${\mathfrak H}=L^2(I_L)\otimes L^2
(\RM)$ the Hamiltonian of the system is formally written as:
\begin{equation}
\label{uzy0}
{\cal H}_{\alpha,L,\omega}\;=\; H^{(p)}\otimes {\mathbb I}\;+\;{\mathbb I}\otimes\hosc\;+\alpha q\;\delta(x)\;,
\end{equation}
where
\begin{equation}
\hosc\;=\;-\;\frac12\frac{\partial^2}{\partial q^2}\;+\;\frac12\omega^2q^2\;.
\end{equation}
is a harmonic oscillator Hamiltonian with frequency $\omega$,
$H^{(p)}_L$ is the Hamiltonian of the particle in the box, and the last term describes the point interaction, which is scaled by the parameter $\alpha>0$.
%is:
%\begin{equation}
%\label{uzy}
%H_{\alpha,L,\omega}\;=\;{\cal H}_{\alpha}\;+\;\hosc\otimes{\mathbb I}\;,
%\end{equation}
%where ${\cal H}_{\alpha}$ is the family of operators which act in ${\mathfrak H}$
%according to $({\cal H}_{\alpha}\psi)(x,q)=H_{\alpha q}\psi(x,q)$, where, for $a$ real,
%\begin{equation}
%\label{hrot}
%H_{a}\;=\;-\frac12\frac{\partial^2}{\partial x^2}\;+\:a\;\delta(x)\;.
%\end{equation}
%with Dirichlet conditions at $x=\pm L/2$. The Dirac function which formally appears in (\ref{hrot}) stands
%for boundary conditions, which connect the left- and the right-hand derivatives in $x=0$ \cite{NS06}\cite{RS2}. \\
 In \cite{smi04} Smilansky's model  was presented in two variants, one with $L=+\infty$ and the other with $L<+\infty$. The spectral theory of the former variant was rigorously analyzed by Solomyak \cite{MS04} and Naboko and Solomyak \cite{NS06}, who proved that for $\alpha>\omega$ a new branch of the absolutely continuous ({\it ac}) spectrum
 of ${\cal H}_{\alpha,\infty,\omega}$ appears besides the one which is naturally associated with unbounded motion of the particle. The new branch coincides with $\RM$, and has  multiplicity $1$.
  The 2nd variant is the  finite-box model which is studied in the present paper. For this variant  an  argument presented in \cite{smi04} shows that normalizable eigenfunctions  exist for $\alpha<\omega$ and don't exist  for $\alpha>\omega$.\\
  %exist for $\alpha<\omega$ but not for $\alpha>\omega$, suggesting  a transition from pure point to continuous  spectrum  at $\alpha=\omega$ .\\
  A generalization of Smilansky's model has $N$ oscillators interacting with the particle at different points. Evans and Solomyak \cite{ES05} have used a scattering theory approach to prove that a spectral transition occurs also in the
 multi-oscillator model with $N=2$ and $L=\infty$ . The nature of their argument makes it intuitively clear , that the same conclusion is true for any $N>0$.\\
    Not much is known about the dynamics of Smilansky's model. In \cite{smi04} strong excitation of the oscillator was surmised for $\alpha>\omega$,  with the particle dwelling near the interaction point. Due to inherent exponential instability of the dynamics, to be proven in the present paper, numerical simulation  of this system is problematic.\\

\subsection{Outline.}
In this paper the dynamics of Smilansky's finite-box model is studied for $\alpha>\omega$, in the
single- and in the multi-oscillator cases (secs. \ref{SOM}, \ref{CWP} respectively). The main results
are Propositions \ref{expeng}, \ref{collaps} and \ref{multicoll}. In the single-oscillator case, the energy of the harmonic oscillator diverges exponentially fast, and, in the limit $t\to\pm\infty$,  the probability distribution of the particle collapses into a $\delta$-function supported in the interaction point.  The dynamical origin of such  exponential instability may be  qualitatively illustrated   as follows: for $q<0$, the $q\delta$ term in the Hamiltonian  acts like
 a potential well for the particle. When $\alpha>\omega$, interaction drives the oscillator still farther in the $q<0$ region. This makes the potential well deeper, and so on. Unbounded increase of the oscillator's energy follows, which is balanced by unbounded decrease of the energy of the particle as it falls deeper and deeper into the well.\\
 This picture is further illustrated by an approximate description of the dynamics, based on a band formalism. This is formally a Born-Oppenheimer approximation in which  the particle plays the role of the fast degree of freedom; however it is a long-time asymptotic approximation
rather than an adiabatic one. In this approximation  the oscillator gets an  effective spring constant, which becomes negative when $\alpha>\omega$.\\
The mathematical groundwork  for the exact dynamical results of Propositions \ref{expeng} and \ref{collaps}  is provided by spectral results largely resting on the work of  Naboko and Solomyak, which  are described in sections \ref{acspec} and \ref{specexp}. In particular, existence of {\it ac} spectrum of multiplicity 1 for $\alpha>\omega$ is assumed as a rigorously proven result, because Naboko and Solomyak's proof \cite{NS06} of a new branch of {\it ac} spectrum in the $L=\infty$ case works, with minor modifications, also in the finite box case.  An independent proof of existence of {\it ac} spectrum (though not of its simplicity) is nevertheless  provided here by spectral expansions, which are constructed in section \ref{specexp} using  formal eigenfunctions.
Such eigenfunctions  are studied in Section \ref{acspec}, by adapting a method used in \cite{NS06}, which includes recourse to Birkhoff's theory
about asymptotic expansions of solutions of 2nd order difference equations \cite{El99}.  In addition, new results about smooth dependence of eigenfunctions on energy,  which are necessary  for the purposes of spectral expansion, are  proven (Note \ref{B&A}), elaborating on  the  formulation by Wong and Li \cite{WL92} of Birkhoff's theory. To be noted that,
unlike the $L=\infty$ case, in the finite-box case  some point spectrum  may survive even above the threshold $\alpha=\omega$, in the presence of special symmetries (see section \ref{acspec}); and that, even in the absence of point spectrum, pure absolute continuity  of the spectrum is not proven (nor it was in the $L=\infty$ case).\\
A finite-box model with an arbitrary finite number of oscillators is studied on a somewhat less rigorous level. Validity of the scattering approach which was developed in \cite{ES05} for the $L=\infty$ variant is assumed, so  a spectral transition is again expected and indeed numerical computations of bands provide evidence that at $\alpha=\omega$
the morphology of the lowest bands undergoes a phase transition , which mirrors the  spectral transition.   \\
 Using  the scattering  approach,  the reduced state of the particle is shown to evolve towards a fully incoherent mixture of "position eigenstates". This process looks like  the wave-packet reduction which is associated with a  measurement of position and indeed the multi-oscillator model is surmised to provide a formal model for approximate position measurement, with the oscillators acting like detectors of the particle's position\footnote{Models with point interactions , different from those considered in this paper, have already been used in studies of decoherence (see, {\it e.g.} \cite{CCF07,CCF05}),  as well as  models of particles interacting with oscillators \cite{DAFT08}.  Approaches to decoherence based on scattering have been used {\it e.g.} in \cite{HS03, CCF05}.}.\\

\section{Single oscillator model.}
\label{SOM}
\subsection{Formal eigenfunctions.}
\label{acspec}
For $\lambda>0$ let $\Lambda: \psi(x,q)\mapsto\lambda\psi(\lambda x,\lambda q)$ denote the unitary scaling operator from
$L^2(I_L)\otimes L^2(\RM)$ to $L^2(I_{\lambda^{-1}L})\otimes L^2(\RM)$. Then:
\begin{equation}
\label{scaling}
\Lambda^{-1}\;{\cal H}_{\alpha,L,\omega}\;\Lambda=\;\lambda^2\;{\cal H}_{\alpha',L',\omega'}\;,
\end{equation}
where $\alpha'=\alpha/\lambda^2$, $L'=L\lambda$, and $\omega'=\omega/\lambda^2$. Therefore,
one of the parameters $L,\omega,\alpha$ may always  be re-set to a prescribed value, by suitably rescaling the coordinates $x$ and $q$ and the time $t$.
Here $L=2\pi$ is assumed, and periodic boundary conditions at $x=\pm\pi$ are used, so the particle may be thought to move in a circle $\mathbb S$ with a distinguished point $O$. This choice affords some formal simplifications without hindering theoretical analysis (see footnote on page 10). That being said,
$\alpha,L,\omega$ will no longer be specified in subscripts  to $\cal H$, unless  strictly necessary.
 As the Hamiltonian is invariant under reflection ($x\mapsto-x$) in the interaction point $O$ , odd functions with respect to $x$ make an invariant subspace. Such functions vanish at the interaction point, so in this subspace
the particle and the oscillator do not interact. For this reason, analysis will be restricted to the invariant subspace  ${\mathfrak H}_+= L^2_+(\SM)\otimes L^2(\RM)$ where $L^2_+(\SM)$ are the square-integrable functions on $\SM$ which are invariant under reflection in $O$.\\
The theory which was developed by Solomyak and Naboko in papers \cite{MS04},\cite{NS06} for the case $L=\infty$ works, with minor  modifications, in the present case as well.
It starts by analyzing the "formal eigenfunctions" of $\cal H$, and this analysis will now be adapted to the present case
because these very eigenfunctions will provide a key to the dynamical analysis to be presented in the following Sections. The wave function $\psi(x,q)$ is expanded
over the normalized eigenfunctions $h_n(q)$ of the harmonic oscillator (Hermite functions):
\begin{equation}
\label{herexp}
\psi(x,q)\;=\;\sum\limits_{n=0}^{\infty}\psi_n(x)\;h_n(q)\;,
\end{equation}
The Hilbert space $\mathfrak H$ is thereby identified with $\ell^2(\NM)\otimes L^2({\mathbb S})$, that is the Hilbert space of sequences $\psi\equiv\{\psi_n(x)\}$ such that $\lgn \psi\rgn^2\equiv\sum_n\|\psi_n\|^2<+\infty$ where $\|.\|$ denotes the $L^2({\mathbb S})$ norm, and the boldface symbol $\lgn.\rgn$ denotes the $\mathfrak H$-norm.
  The Hamiltonian is formally identified with
the differential operator which acts in $\tp$ according to:
\begin{gather}
\{\psi_n(x)\}\;\;\mapsto\;\;\{L_n\psi_n(x)\}\;,\nonumber\\
L_n\;=\;-\frac12\frac{d^2}{dx^2}\;+\;\biggl(n+\frac12\biggr)\omega\;,\;\;\;\;(\;n=0,1,2,\ldots\;)\;.
\label{ellenne}
\end{gather}
A vector $\psi\in\ell^2(\NM)\otimes L^2({\mathbb S})$ is in the domain of the Hamiltonian if  each $\psi_n$ has  a
square-integrable 2nd derivative in $\mathbb{S}\setminus\{O\}$, such that  $\sum_n\|L_n\psi_n\|^2<+\infty$, and, moreover,
certain boundary conditions  at $x=0$ are  satisfied. These  are  dictated by the $\delta$ function in (\ref{uzy0}).  Using recurrence properties of the Hermite functions, such "matching conditions"  may be written in the form \cite{NS06}:
\begin{equation}
\label{flux}
\psi'_n(0+)\;-\psi'_n(0-)\;=\;2\alpha\;(2\omega)^{-1/2}\bigl( \sqrt{n+1}\;\psi_{n+1}(0)\;+\;\sqrt{n}\;\psi_{n-1}(0)\bigr)\;,
\end{equation}
where the rightmost term is $0$ for $n=0$. The "formal eigenfunctions"  are sequences $\{u_n(x,E)\}$  that solve the infinite system of equations
\begin{equation}
\label{infsy}
L_n\;u_n(x,E)\;=\;E\;u_n(x,E)\;,\;\;\;n=0,1,\ldots,
\end{equation}
 for $E\in\RM$,  and satisfy the matching conditions (\ref{flux}). Such sequences need not belong in $\ell^2(\NM)\otimes L^2({\mathbb S})$, and are sought in the form:
 \begin{equation}
 \label{defC}
 u_{n}(x,E)\;=\;C(n,E)\;v_n(x,E)\;,
 \end{equation}
where, for each $n=0,1,2,\ldots$, the functions:
\begin{equation}
\label{geneigf1}
v_n(x,E)\;=\;\rho_n(E)\;\cos\;\bigl(k_n(E)(|x|-\pi)\bigr)\;\;,\;\;\;k_n(E)=\sqrt {2E-(2n+1)\omega}\;
\end{equation}
 are normalized solutions of eqn.(\ref{infsy}). The factor $\rho_n$ in (\ref{geneigf1}) is chosen so that $\|v_n(.,E)\|=1$ :
\begin{equation}
\label{geneigf2}
\rho_n(E)\;=\;\biggl(\pi+\frac{\sin(2k_n(E)\pi)}{2k_n(E)}\biggr)^{-1/2}\;.
\end{equation}
For $n>E/\omega -1/2$,~ $k_n(E)$ is imaginary and  formulae (\ref{geneigf1}), (\ref{geneigf2}) are conveniently  rewritten with circular functions replaced by hyperbolic ones, and $k_n(E)$ replaced by $\chi_n(E)\equiv-ik_n(E)$. \\
Coefficients $C(n,E)$ in (\ref{defC}) have to be chosen such that conditions (\ref{flux}) be satisfied. Substituting (\ref{defC}) and (\ref{geneigf1}) in eqs. (\ref{flux}) one finds that to this end they must solve the following 2nd order difference equation :
\begin{gather}
 \label{bkad1}
 h_2(n,E)\;C(n+2,E)\;+\;h_1(n,E)\;C(n+1,E)\;+\;h_0(n,E)\;C(n,E)\;=\;0\;,\;\;\;\;\;(n\geq 0)\;,
  \end{gather}
with  initial conditions
 $C(0,E)$ and $C(1,E)$ that satisfy:
 \begin{equation}
 \label{condin}
h_2(-1,E)\;C(1,E)\;=\;-h_1(-1,E)\; C(0,E)\;,
\end{equation}
 having denoted, for $n\geq -1$ :
 \begin{gather}
 \label{p0p1}
 h_2(n,E)\;=\;\alpha\;\sqrt{n+2}\;\;v_{n+2}(0,E)\;;\;\;\;h_1(n,E)\;=\;(2\omega)^{1/2}\;v'_{n+1}(0+,E)\;,
 \end{gather}
 and for $n\geq 0$:
 \begin{gather}
 h_0(n,E)\;=\;\alpha\;\sqrt{n+1}\;v_n(0,E)\;.
 \label{p1p2}
 \end{gather}
 Let ${\cal E}_{\omega}$ denote the set of real energies $E$ such that either $h_2(n,E)$ or $h_0(n,E)$ or both vanish  for some $n\geq 0$:  such energies   are given by $2E=(2n+1)\omega+(r+1/2)^2$  for  $r\in\ZM$ and $n\geq 0$, so  ${\cal E}_{\omega}$ has at most finite intersection with any bounded interval.
 Whenever $E\in\RM\setminus{\cal E}_{\omega} $,  solutions of eqn. (\ref{bkad1}) are in one-to-one correspondence with their initial values $C(0,E)$ and  $C(1,E)$. In particular,  the  solution of (\ref{bkad1}) which verifies (\ref{condin}) exists,  and is uniquely fixed by the value of $C(0,E)$, which plays the role of a normalization constant.
An asymptotic approximation as $n\to+\infty$ to this solution is provided by a theory of Birkhoff and Adams \cite{El99}, which is briefly reviewed in Note \ref{B&A}. According to that theory,  if $\alpha>\omega$ and $E\in\RM\setminus{\cal E}_{\omega}$ then equation (\ref{bkad1}) has two "normal" linearly independent solutions $C^{\star}_{\pm}(n,E)$ which have the following $n\to +\infty$ asymptotics:
 \begin{equation}
 \label{bkad2}
 C_{\pm}^{\star}(n,E)\;\sim\;\frac{1}{\sqrt{n}}\;\exp (\pm in\theta\;\mp i\lambda E\log(n))\;\;+O(n^{-3/2})\;,
 \end{equation}
 where
\begin{equation}
\label{bkad77}
\theta\;=\;\arccos(\omega/\alpha)\;,\;\;\;\;\lambda\;=\;\frac1{2\sqrt{\alpha^2-\omega^2}}\;.
\end{equation}
(Cp. Lemma 3.3 in Ref. \cite{NS06}, where parameters $\mu$ and $\Lambda$ respectively correspond to $\omega/\alpha$ and $E/\omega$). Thanks to reality of coefficients (\ref{p0p1}) and (\ref{p1p2}), the complex conjugate of any solution of
eqn. (\ref{bkad1}) is still a solution. Hence the normal solutions are mutually conjugate, because such are their asymptotic approximations (\ref{bkad2}). If $C(0,E)$ is chosen real in (\ref{condin}), then the sought for particular solution $C(n,E)$ is  real for all $n>0$, so  it  may  be written as a linear superposition of the mutually conjugate "normal" solutions, in  the following form :
\begin{equation}
\label{nmfct}
C(n,E)\;=\;C(0,E)\;\real\;\bigl\{Z(E)C^{\star}_+(n,E)\bigr\}\;,
\end{equation}
where $Z(E)\in\CM$ has to be chosen so that:
\begin{equation}
\label{zeta}
\real\;\bigl\{Z(E)C^{\star}_+(0,E)\bigr\}\;=\;1\;,\;\;\;\;h_2(-1,E)\;\real\;\bigl\{Z(E)C^{\star}_+(1,E)\bigr\}\;=\;-h_1(-1,E)\;,
\end{equation}
as is required by the initial condition (\ref{condin}). In the rest of this work, the normalization factor  $C(0,E)$ in (\ref{nmfct}) will be fixed such  that  $C(0,E)|Z(E)|=\pi^{-1/2}$. This choice is aimed at Lemma 1 below. Thanks to it,   coefficients $C(n,E)$ of the formal  eigenfunction $\{u_n(x,E)\}$ (cp. (\ref{defC})) have the following asymptotic form:
\begin{equation}
\label{asycn}
C(n,E)\;\sim\;\frac {1}{\sqrt{\pi n}}\;\cos\bigl(n\theta\;-\;\lambda{E}\log(n)\;+\;\zeta(E)\bigr)\;+\;O(n^{-3/2})\;.
\end{equation}
Here $\zeta(E)$ is the phase of $Z(E)$; it is not explicitly known, because  the normal solutions $C_{\pm}^{\star}(n,E)$ are  not  known except for their asymptotic forms,  so eqs. (\ref{zeta}) cannot be solved explicitly. In Note \ref{B&A} (Corollary  \ref{mostimp}) $Z(E)$ is  proven to be a $C^1$ function of $E$ in any closed interval $I$ having empty intersection
with the set ${\cal E}_{\omega}^*$ of ''exceptional''  energies, which is obtained on adding to ${\cal E}_{\omega}$ the threshold energies $(n+1/2)\omega$,
$(n\geq 0)$, which are branch points for  coefficients in eqn.(\ref{bkad1}).

It should be stressed that ({\ref{asycn}) only holds when $\alpha>\omega$, and that the case $\alpha\leq\omega$ is not discussed here.
Of crucial importance is that:
\begin{equation}
\label{veryimp}
\sum\limits_{n=0}^{+\infty}|C(n,E)|^2\;=\;+\infty\;,
\end{equation}
so the sequence $\{u_{n}(x,E)\}$ is not in $\ell^2(\NM)\otimes L^2(\mathbb{S})$ and does not define an eigenvector proper.
It is worth noting, however, that the series $\sum_nC(n,E)v_n(x,E)h_n(q)$ is pointwise convergent to a
$C^{\infty}$ function $\varphi_E(x,q)$ at all points
 $(x,q)\in({\mathbb S}\setminus\{0\})\times\RM$, because if $|x|>\delta>0$ then $v_n(x,E)$ decays quite fast,
$\sim 2n^{1/4}e^{-\delta\sqrt{2n}}$ with $n$, and
the Hermite functions are uniformly bounded \cite{EMOT55}.\\
Such properties of the infinite recursion (\ref{bkad1}) are essentially identical to
ones which were established in \cite{NS06} for the $L+\infty$ case. Due to them,
for $\alpha>\omega$ the spectrum of ${\cal H}_{\alpha,L,\omega}$ acquires an absolutely continuous component, with multiplicity $1$.

\subsection{Spectral expansions.}
\label{specexp}

Throughout the following, $\alpha>\omega$ is understood, and the absolutely continuous subspace of $\cal H$ is  denoted ${\mathfrak H}_{\text{\tiny ac}}$. In this Section the above described formal eigenfunctions $\{u_n(x,E)\}$ are used to construct spectral expansions.

\begin{lemma}
\label{normeig}
{If $\Psi,\Phi\in C_0(\RM\setminus{\cal E}_{\omega}^*)$ (the continuous, compactly supported functions having no exceptional energies in their support),  then
\begin{equation}
\label{orth}
\sum\limits_{n=0}^{\infty}\iint\limits_{\RM^2}dE_1\;dE_2\;{\bar \Psi}(E_1)\Phi(E_2)\int_{\SM}dx\;u_n(x,E_1)u_n(x,E_2)\;=\;
\int_{\RM}dE\;{\bar \Psi}(E)\Phi(E)\;.
\end{equation}
}
\end{lemma}
{\it Proof}:
let $P_n(E_1,E_2)=(u_n(E_1),u_n(E_2))$ denote the scalar product in $L^2_+(\SM)$ of $u_n(x,E_1)$ and $u_n(x,E_2)$. From
 $L_nu_n=Eu_n$  (cp. eqn.(\ref{ellenne})):
 \begin{gather}
 P_n(E_1,E_2)\;=\;-(2E_1)^{-1}\biggl(\int_0^{\pi}\;+\;\int_{-\pi}^0\;\biggr)dx\;
 u_n(x,E_2)\frac{\partial^2}{\partial x^2}\;u_n(x,E_1)\;+\nonumber\\
 +\;(2E_1)^{-1}(2n+1)\bigl(u_n(E_1),u_n(E_2)\bigr)\;.
 \end{gather}
 Integration by parts yields:
 \begin{gather}
 \label{orth2}
 2(E_1\;-\;E_2)\;P_n(E_1,E_2)\;=\
 \bigl[
 u'_n(0+,E_1)-u'_n(0-,E_1)
 \bigr] u_n(0,E_2)\;+\nonumber\\
 -\;
 \bigl[u'_n(0+,E_2)-u'_n(0-,E_2)
 \bigr] u_n(0,E_1)
 \;.
 \end{gather}
 Using the matching condition (\ref{flux}),
 \begin{gather}
 \label{orth3}
 (E_1\;-\;E_2)\;P_n(E_1,E_2)\;=\;\frac{\alpha}{\sqrt{2\omega}}\;\bigl(W_{n+1}(E_1,E_2)\;-\;W_n(E_1,E_2)\bigr)\;,
 \end{gather}
 where, for $n>0$
 \begin{equation}
 \label{oorth}
 W_n(E_1,E_2)\;=\;\sqrt{n}\;\bigl(u_n(0,E_1)u_{n-1}(0,E_2)\;-u_{n-1}(0,E_1)u_n(0,E_2)\bigr)
 \end{equation}
 and $W_0(E_1,E_2)=0$. Hence,
 $$
 \sum\limits_{n=0}^NP_n(E_1,E_2)\;=\;\frac{\alpha}{\sqrt{2\omega}}\;\frac{W_{N+1}(E_1,E_2)}{E_1-E_2}\;.
 $$
Substituting  in (\ref{oorth}) the asymptotic form of $u_n(0)$ which follows from eqs.(\ref{geneigf1}),(\ref{geneigf2}), (\ref{asycn}):
\begin{gather}
 \sum\limits_{n=0}^NP_n(E_1,E_2)\;\sim\;
 {2\alpha\sin(\theta)}\;\frac{\sin\bigl((\lambda(E_1-E_2)\ln(N)-\zeta(E_1)+\zeta(E_2)\bigr)}{\pi(E_1-E_2)}\nonumber\\
 \underset{N\to\infty}{\longrightarrow}\;\;\delta(E_1-E_2)\;.
\end{gather}
where the definitions (\ref{bkad77}) of $\lambda$ and $\theta$ have been used. Convergence to the Dirac delta function is meant in the sense of eqn.(\ref{orth}) for continuous
functions $\Psi$ and $\Phi$ supported in $\RM\setminus{\cal E}^*_{\omega}$  and rests
 on regularity of $\zeta(E)$ in $\RM\setminus{\cal E}^*_{\omega}$, as established by Corollary \ref{mostimp} in Note \ref{B&A}.$\Box$

Thanks to (\ref{orth}),  whenever $\Psi\in C_0(\RM\setminus{\cal E}^*_{\omega})$ the sequence of functions which are defined on $\SM$ by
\begin{equation}
\label{cint}
\psi_n(x)\;=\;\int_{\RM}dE\;\Psi(E)u_n(x,E)\;,\;\;\;(n=0,1,2,\ldots)
\end{equation}
is a vector in $\tp$. This vector will be denoted $\psi$, and the function $\Psi$ will be termed the spectral representative of $\psi$.
Eqn.(\ref{orth}) says that the map $\imath:\Psi\mapsto\psi$ is isometric, so $\imath$ extends to an isometry of $L^2(\RM)$ into $\tp$ . Proposition \ref{specmap} below easily follows . It  shows that this  map, or rather its inverse, yields a complete spectral representation of ${\cal H}$ restricted to its absolutely continuous subspace.
\begin{proposi}
\label{specmap}
{;\\\
(i) For all $\Psi\in L^2(\RM)$, and $t\in\RM$,
\begin{equation}
\label{specth}
\imath \bigl(e^{-iEt}\Psi\bigr)\;=\;e^{-i{\cal H}t}\imath(\Psi)\;,
\end{equation}
(ii) $\imath$ is a unitary isomorphism of $L^2(\RM)$ onto ${\mathfrak H}_{\text{\tiny ac}}$.\\
}
\end{proposi}
A proof is presented in Note \ref{proofspecmap}.

\subsection{Dynamics for $\alpha>\omega$.}
 The spectral results in the previous Sections  have dynamical consequences as stated in the Propositions below. Throughout this section  $\alpha>\omega$ is understood.
  Let $\psi\in L^2_+(\SM)\otimes L^2(\RM)$ , and $\|\psi\|=1$. The notation $\psi(t)=e^{-i{\cal H}t}\psi$ will be used; moreover, $\langle f\rangle_T=\tfrac1T\int_0^Tdtf(t)$ will denote the time average up to time $T$ of a function $f(t)$.

\begin{proposi}
\label{expeng}
{If $\psi\in {\mathfrak H}_{\text{\tiny ac}}$ then the time-averaged  energy of the oscillator  grows in time, at least exponentially fast: i.e.,
\begin{equation}
\label{expgrth}
\liminf\limits_{T\to+\infty}\;\frac{\ln(E_{\mbox{\rm\tiny osc}}(T))}{T}\;>\;0\;.
\end{equation}
where
$$
E_{\mbox{\rm\tiny osc}}(T)\;=\;\biggl\langle\;\bigl(\psi(t)\;,
\;{\mathbb I}\otimes\hosc\;\psi(t)\bigr)\;\biggr\rangle_T\;.
$$
}
\end{proposi}
A proof is given in Appendix \ref{app1}.
\begin{proposi}
\label{collaps}
{If $\psi\in {\mathfrak H}_{\text{\tiny ac}}$ and $\|\psi\|=1$ then the probability distribution of the  position $x$ of the particle weakly converges to $\delta(x)$ in the limit
$t\to+\infty$.}
\end{proposi}
This  is equivalent to
\begin{equation}
\label{delta}
\lim\limits_{t\to+\infty}\int dq\int\limits_{\eta<|x|<\pi}dx\;|\psi(q,x,t)|^2\;=\;0
\end{equation}
for any $0<\eta<\pi$, which is proven in Appendix \ref{app2}.\\
 One may reasonably expect the expectation  value of the position $q$ of the oscillator to diverge to $-\infty$ as the particle endlessly falls in the $\delta$-potential . This will  be further supported by arguments in the next Section; however no exact proof is attempted here.\\
The reduced state of the particle when the full system is in the pure state $\psi(t)$ is the positive trace class operator $S_{\psi}(t)$ in $L^2(\SM)$ such that $\Tr(S_{\psi}(t)A)=(\psi(t), A\otimes{\mathbb I} \psi(t))$ for all bounded operators $A$ in $L^2(\SM)$. Proposition \ref{collaps} entails a somewhat extreme form of decoherence for the  reduced state:
\begin{coro}
\label{dec}
{If $\psi\in {\mathfrak H}_{\text{\tiny ac}}$ then for every $\phi$ and $\phi'$ in $L^2(\SM)$,
$
\lim_{t\to\pm\infty}(\phi, S_{\psi}(t)\phi')=0$ . }
\end{coro}
A proof is presented in  Note \ref{proofcor}.

\begin{figure}
\includegraphics[width=8cm,angle=0]{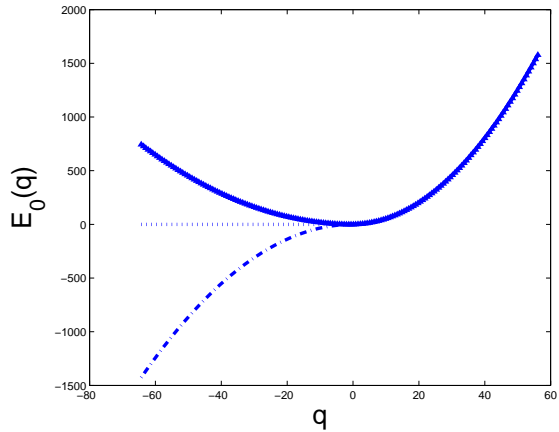}
\caption{The lowest energy band in the single oscillator model for $\omega=1$ and $\alpha=0.8$ (upper, full line),
$\alpha=1$ (middle, dotted line) and $\alpha=1.3$ (lower, dashed-dotted line).
}
\label{pacfig1}
\end{figure}

\subsection{Band dynamics: inverted oscillator.}
\label{bandyn}
An intuitive picture of the above results is provided by an approximate description of the dynamics, to be presented in this section. Hamiltonian (\ref{uzy0}) (with $L=2\pi$) may also be presented in the following form:
\begin{equation}
\label{uzyad}
{\cal H}\;=\;\int^{\oplus
}_{\RM}dq\;\mathrm{H}_{\alpha}(q)\;+\;{\mathbb I}\otimes\hosc\;,
\end{equation}
where, for any fixed value of $q$,
\begin{equation}
\label{accaq}
\mathrm{H}_{\alpha}(q)\;=\; H^{(p)}\;+\;\alpha q\;\delta(x)
\end{equation}
is an operator in $L^2_+(\SM)$ \cite{RS2}. It has a complete set of eigenfunctions and eigenvalues which parametrically depend on the product $\alpha q$. All eigenfunctions are real valued and have the form:
\begin{equation}
\label{bdegf}
\phi_{q,n}(x)\;=\;A_n\;\cos(\xi_n(|x|-\pi))\;\;\,\;\;n=0,1,2,\ldots
\end{equation}
where $A_n$ are normalization constants, and $\xi_n$ are the solutions  with $\Re{\text e}(\xi)\geq 0$ of the equation:
$$
\tan(\pi\xi)\;=\;\frac{\alpha q}{\xi}\;.
$$
They are numbered in increasing order of the corresponding  energy eigenvalues  $W_n(q)=\tfrac12 \xi_n^2$. All  $\xi_n$ with $n>0$ are real,  and behave like
$\xi_n\sim n+\tfrac12$ asymptotically as $q\to \pm\infty$. Instead $\xi_0$ is  real only when $q>0$, and turns imaginary
when $q<0$ ; in that case  $\xi_0=i\chi$, where $\chi$ is the unique positive solution of
$$
\tanh(\pi\chi)\;=\;-\frac{\alpha q}{\chi}\;.
$$
Hence $W_0(q)$ is negative whenever $q<0$. A standard Feynman-Hellman argument yields:
\begin{equation}
\label{adia2}
\frac{dW_n(q)}{dq}\;=\;\alpha\;\phi_{q,n}(0)^2\;,
\end{equation}
so the levels $W_n(q)$ are nondecreasing functions of $q$. The ground state energy $W_0(q)$ is asymptotically given by:
\begin{equation}
\label{gren}
W_0(q)\;=\;\tfrac12\xi_0^2\;\sim \left\{
             \begin{array}{ll}
               \tfrac18, & \hbox{for\;$q\to+\infty$\;;} \\
               -\tfrac12\alpha^2q^2, & \hbox{for\;$q\to -\infty$\;.}
             \end{array}
           \right.
\end{equation}
When $q>0$, the ground state eigenfunction has still the form (\ref{bdegf}) with $n=0$.  For $q<0$ it is instead given by:
$$
\phi_{q,0}(x)\;=\;A_0\;\cosh\bigl(\chi(q)(|x|-\pi)\bigr)\;,\;\;\;(q<0)\;.
$$
so for large negative $q$ it is sharply peaked at $x=0$.
 Any $\psi\in L^2_{+}(\SM)\otimes L^2(\RM)$
may be expanded as
\begin{equation}
\label{quq}
\psi(x,q)\;=\;\sum\limits_{n=0}^{\infty}Q_n(q)\;\phi_{q,n}(x)\;\;,\;\;\;Q_n(q)=\int_{-\pi}^{\pi}dx\;\phi_{q,n}(x)\psi(x,q)
\;,
\end{equation}
so that
$$
\sum\limits_{n=0}^{\infty}|Q_n(q)|^2\;=\;\int_{-\pi}^{\pi}dx\;|\psi(x,q)|^2\;.
$$
In this way $\mathfrak H$ is decomposed in "Band Subspaces" ${\mathfrak B}_n\equiv\{
\psi(x,q)=Q_n(q)\phi_{q,n}(x)\;|\;Q_n(q)\in L^2(\RM)\}$. The projector onto the $n$-th band subspace will be denoted $\Pi_n$.
In the "band  formalism" it is  easy to show that ${\cal H}_{\alpha,\omega}$ is bounded from below when $\alpha<\omega$. Indeed, if $\psi$ is in the domain of ${\cal H}_{\alpha,\omega}$, then
\begin{eqnarray}
(\psi,{\cal H}_{\alpha,\omega}\psi)\;&=&\;\sum\limits_{n=0}^{\infty}\int_{-\infty}^{\infty}dq\;W_n(q)|Q_n(q)|^2\;+\;(\psi,{\mathbb I}\otimes \hosc\psi)\;.
\end{eqnarray}
Singling out the contribution of the lowest band $\mathfrak B_0$, and using (\ref{gren}) and monotonicity of $W_0(q)$:
\begin{eqnarray}
(\psi,{\cal H}_{\alpha,\omega}\psi)&\geq&\;-\frac12\alpha^2\int_{-\infty}^{\infty}dq\;\int_{-\pi}^{\pi}dx\;q^2|\psi(x,q)|^2\;+\nonumber\\
&\;&\;+\;\sum\limits_{n=1}^{\infty}\int_{-\infty}^{\infty}dq\;W_n(q)|Q_n(q)|^2\;+
\;(\psi,{\mathbb I}\otimes\hosc\;\psi)\nonumber\\
&\geq&\;(\psi,{\mathbb I}\otimes
H^{\mbox{\rm\tiny (osc)}}_ {\sqrt{\omega^2-\alpha^2}}\psi)\nonumber\\
&\geq&\;\frac12\sqrt{\omega^2-\alpha^2}\;\|\psi\|^2\;.
\end{eqnarray}
This shows that the abrupt  change from semibounded to unbounded spectrum which occurs at $\alpha=\omega$ is related to
a change in the structure  of the
"ground band" alone. This transition is further elucidated by noting that the ground band  plays the role of a stable variety, due to the following:
\begin{proposi}
\label{BO}
{If $\alpha>\omega$ then, for arbitrary $\psi\in{\mathfrak H}_{\mbox{\tiny ac}}$, and for all integer $n>0$:
$$
\lim\limits_{t\to\infty}\|\;\Pi_n e^{-i{\cal H}_{\alpha,\omega}t}\psi\;\|\;=\;0\;.
$$
}\end{proposi}
A proof is presented  in section \ref{BOproof}.
This suggests that asymptotic solutions in time of the time-dependent Schr\"odinger equation may be sought in the form
$\psi(x,q,t)=Q_0(q,t)\phi_{q,0}(x)$. No exact proof  is attempted here; nevertheless, direct substitution yields a Schr\"odinger equation for the
band wavefunction $Q_0(q,t)$:
$$
i\frac{\partial Q_0}{\partial t}\;=\;-\frac12\frac{\partial^2 Q_0}{\partial q^2}\;+\;{\cal V}(q)Q_0\;,
$$
where the band potential ${\cal V}(q)$ is given by:
\begin{equation}
\label{bandpt}
{\cal V}(q)\;=\;\frac12\omega^2 q^2\;+\;W_0(q)\;+\;\frac12\int_{\SM}dx\;\biggl|\frac{\partial\phi_{q,0}(x)}{\partial q}\biggr|^2\;.
\end{equation}
A calculation reported in Note \ref{boextra} shows that the last term on the right-hand side is $O(q^{-2})$
as $q\to-\infty$ and tends to a constant when $q\to+\infty$. At large $q$, the band potential is then determined by the other two terms. The band potential which results  of these two terms alone  is shown in Fig.1. As $\alpha$ grows beyond $\omega$, it turns from  concave to monotone increasing and then, at large negative $q$,
${\cal V}(q)\sim -\frac12(\alpha^2-\omega^2)q^2$, that is the potential of an inverted harmonic oscillator. This
sort of phase transition qualitatively explains the  growth of the harmonic oscillator's energy which was proven in Proposition 1, and suggests that it is exponential with rate $\lambda^{-1}=2\sqrt{\alpha^2-\omega^2}$.
\section{Collapse of wave-packets.}
\label{CWP}
\subsection{A multi-oscillator model.}
\label{moscmod}

 Smilansky's model has generalizations, in which  an arbitrary  finite number $N$ of harmonic oscillators interact with the particle at different points $O_1,\ldots,O_N$ . In the $L=\infty$ case, the corresponding spectral theory has been developed by Evans and Solomyak \cite{ES05} using a  scattering theory approach. Translated to the present case, this approach is as follows.
 All oscillators are assumed to have the same frequency $\omega$ and coupling constant $\alpha$, and  a circular ordering is assumed for
 the interaction points $O_i$.  For each $i=1,\ldots,N$ let a rigid wall be inserted
 at a point $Z_i$  in between  $O_i$ ad $O_{i+1}$, and let $J_i$ denote the arc
 $Z_i,Z_{i+1}$. The  Hamiltonian ${\cal H}^{(b)}$ of the resulting system differs from ${\cal H}$ because of  Dirichlet conditions at the points $Z_i$, and is actually an orthogonal sum of operators ${\cal H}^{(b)}_i$ in $
 {\mathfrak H}_i\equiv L^2(J_i)\otimes L^2(\RM^N)$, each of which describes the particle in a rigid box $J_i$, coupled to the $i$-th oscillator alone. Therefore, thanks to what is known about the single-oscillator box model,
 ${\cal H}^{(b)}$ has {\it ac} spectrum coinciding with $\RM$ when  $\alpha>\omega$.
M{\o}ller wave operators are defined by :
 \begin{equation}
 \label{moll}
\Omega_{\pm}({\cal H},{\cal H}^{(b)})\;=\;\lim\limits_{t\to\pm\infty}e^{i{\cal H}t}\;e^{-i{\cal H}^{(b)}t}\;P_{ac}^{(b)}
 \end{equation}
where $P_{ac}^{(b)}$ denotes projection onto the absolutely continuous subspace of ${\cal H}^{(b)}$. They are said to be complete if their range coincides with the entire absolutely continuous subspace of
${\cal H}$. Whenever this happens, the wave operators $\Omega_{\pm}({\cal H}^{(b)},{\cal H})$ also exist \cite{RS3}.
Existence and completeness have been proven in \cite{ES05} for the model with $L=\infty$.
Here they are assumed also for the multi-oscillator box model. In the absence of a formal proof paraphrasing Evans and Solomyak's, this assumption rests on intuition provided by Proposition 2:  as the wave function which evolves in ${\mathfrak
 H}_i$ under Hamiltonian ${\cal H}^{(b)}_i$ is drained by the $i$-th interaction point,  the boundaries  at $Z_{i}$ and $Z_{i+1}$ become ininfluent, and so does the difference between $\cal H$ and ${\cal H}^{(b)}$ .\footnote{ The same picture accounts for
irrelevance of boundary conditions in the single-oscillator box model.}  \\
Existence and completeness of wave operators enforce unitary equivalence of the absolutely continuous parts of $\cal H$ and of ${\cal H}^{(b)}$, hence  infinitely degenerate
Lebesgue  spectrum of $\cal H$ at $\alpha>\omega$.
\subsection{Band formalism.}

There is a band formalism also for the N-oscillators case.  The case $N=2$ will be briefly described. Oscillators $1$ and $2$ with respective coordinates $q_1$ and $q_2$ are coupled to the particle at points  $x=0$ and $x=\pm\pi$ respectively,  diametrally opposite in $\SM$. The particle Hamiltonian which now replaces (\ref{accaq}) parametrically depends on $q_1$ and $q_2$, and has real-valued
 eigenfunctions $\phi_{q_1,q_2,n}(x)$ in $L^2(\SM_+)$  and eigenvalues $W_n(q_1,q_2)=\tfrac12\xi_n^2$,  where $\xi_n$ are the solutions with $\Re{\text e}(\xi_n)\geq 0$ (numbered in non-decreasing order of the corresponding eigenvalues) ) of the equation:
 \begin{equation}
 \label{seceq}
 \tan(\pi\xi)\;=\;\frac{\alpha\xi(q_1+q_2)}{\xi^2-\alpha^2q_1q_2}.
 \end{equation}
 "Band potentials"  $E_n(q_1,q_2)=\tfrac12\omega^2(q_1^2+q_2^2)+W_n(q_1,q_2)$ computed by numerically solving eq. (\ref{seceq}) are shown in Fig.2 and in Fig.3. Like in the $N=1$ case,  the spectral transition at $\alpha=\omega$ is  concomitant to  a phase transition in the structure of the lowest band.  A transition is observed  for the 2nd lowest energy band as well, at a higher value of $\alpha/\omega\approx 1.414$ (not shown), but not for higher bands. The structure of  the overcritical ground band, shown in Fig. 3, is explained as follows.
When $1<\alpha/\omega$ , $\xi_0$ is found to be imaginary in the region $\mathcal R$ of the $(q_1,q_2)$ plane  which is defined by the inequality:
 $$
 q_-\;<\;-\frac{q_+}{1+\pi q_+}\;,
 $$
 where $q_-$ and  $q_+$ are respectively  the minimum and  the maximum of $q_1$ and $q_2$. Moving out to infinity in $\cal R$ along a half-line started at $(0,0)$,  the asymptotic behaviour of $\xi_0$ is $\sim -i\alpha q_-$, so the band potential $E_0(q_1,q_2)$ diverges  to $-\infty$, except along directions lying within an angle of $\arcsin(\omega/\alpha)-\pi/4$
 on either side of the half-line $q_1=q_2<0$, where instead  it diverges to $+\infty$ as long as $\omega<\alpha<\omega\sqrt{2}$. This is why in Fig.3 one observes two valleys , hereby labeled $1$ and $2$, that descend to $-\infty$ along  the negative $q_1$ axis and the negative $q_2$ axis respectively, and  are separated by a crest, which rises along  the $q_1=q_2<0$ half-line.
 In region  $\mathcal R$  the ground eigenfunction has the form
 $$
 \phi_{q_1,q_2,0}(x)\;=\;C_1(q_1,q_2)\;e^{|\xi_0(q_1,q_2)||x|}\;+\;C_2(q_1,q_2)e^{-|\xi
 _0(q_1,q_2)||x|}\;.
 $$
 It has two peaks, labeled  $1$ and $2$,  at the interaction points of oscillators $1$ and $2$.  Descending along either valley both peaks become narrower and narrower, however calculation shows that the whole probability is asymptotically in time  caught within  the peak which shares  the label of the valley.\\
Like in sec. \ref{bandyn}, for $\omega<\alpha<\omega\sqrt{2}$, the quantum dynamics is asymptotically attracted by the $0$-th band subspace,  which consists  of functions of the form $Q_0(q_1,q_2)\phi_{0,q_1,q_2}(x)$. The "band wave-function" $Q_0(q_1,q_2,t)$ asymptotically in time  solves the Schr\"odinger equation for a particle in the plane $(q_1,q_2)$, subject to a potential that behaves at $\infty$ like the one in Fig. 3.
  A classical particle would escape to infinity along one valley, so one and just one oscillator would undergo unbounded excitation.
\begin{figure}
\includegraphics[width=8cm,angle=0]{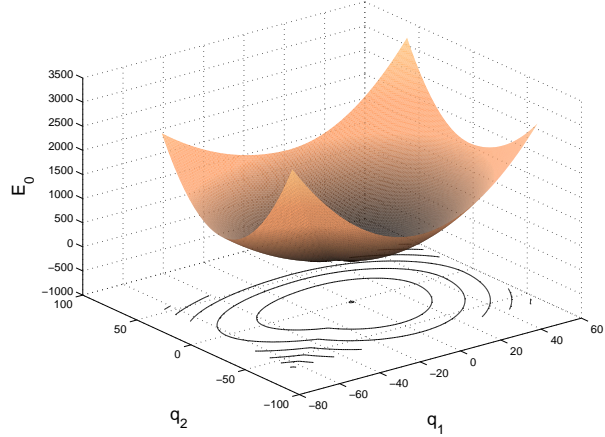}
\caption{ The lowest energy band in the model with 2 oscillators, for $\alpha=0.7$ and $\omega=1$.
}
\label{pacfig2}
\end{figure}
\begin{figure}
\includegraphics[width=12
cm,angle=0]{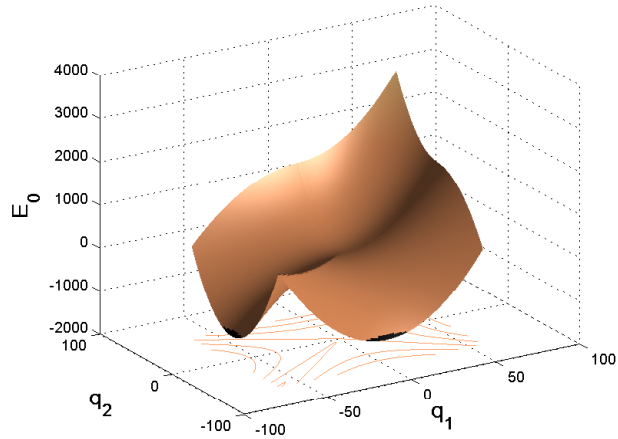}
\caption{Same as Fig.\ref{pacfig2},
for $\alpha=1.3$ and $\omega=1$.
}
\label{pacfig3}
\end{figure}

\subsection{Collapse of wave-packets.}
Existence and completeness of wave operators have the following immediate consequence, which generalizes Proposition
\ref{collaps}:
\begin{proposi}
\label{multicoll}
{If $\alpha>\omega$ and $\psi\in{\cal H}_{ac}$ with $\|\psi\|=1$, then  the probability distribution
of the particle converges weakly as $t\to\pm\infty$ to a superposition of $\delta$ functions supported in the interaction points.
$$
\idotsint\limits_{\RM^N}dq_1\ldots dq_N\;|\psi(x,q_1,\ldots,q_N,t)|^2
\;\;\underset{t\to\pm\infty}\longrightarrow\;\;\sum\limits_{j=1}^N\gamma_j^{\pm}\;\delta(x-O_j)\;,
$$
 where:
\begin{equation}
\label{weights}
 \gamma_j^{\pm}\;=\;\|P_j\;\Omega_{\pm}({\cal H}^{(b)},{\cal H})\psi\|^2\;,
\end{equation}
and $P_j$ denotes projection onto ${\mathfrak H}_j$.}
\end{proposi}
A proof is presented in section \ref{proofmulticoll}. This in particular implies that the right-hand sides in (\ref{weights}) do not depend on the positions $Z_i$ of the rigid walls.\\
Corollary \ref{dec} about complete decoherence of the reduced state of the particle  generalizes to the multi-oscillator case. Hence one may say that
coupling to the oscillators causes the reduced state of the particle to evolve exponentially fast towards  an "incoherent superposition of position eigenstates". In the case when the particle is initially in a pure state,
this process is similar to the  wave-packet reduction which is conventionally associated with measurements of position.
Here the measuring apparatus consists  of $N$ oscillators, and the $N$ interaction points have to be  chosen in   a  thick homogeneous grid.
 Under the assumptions in Proposition \ref{multicoll}, as $t\to+\infty$  the pure state $\psi(t)=e^{-i{\cal H}t}\psi$ comes closer and closer to the state $e^{-i{\cal H}^{(b)}t}\Omega^+\psi$. This  is a coherent superposition of states which have the particle in a box and the corresponding oscillator in a highly excited state.
 Tracing out the oscillators yields an incoherent mixture of alternatives for the particle position , however the probability of finding the particle in the $m$-th box  is
 given by $\gamma^+_m=\|P_m\Omega_+\psi\|^2$ as in eqn. (\ref{weights}) and not by   $\int_{J_m}dx|\psi_0(x)|^2=\|P_m\psi\|^2$ as in an ideal measurement. The difference lies with replacing the initial state $\psi$ with the "outgoing state" $\Omega_+\psi$ , and may be ascribed to the non-instantaneous nature of the measurement process.
  Increasing the number $N$ of  oscillators (hence increasing the precision of the measurement)
  while keeping $\alpha/N$ and $\omega/N$ constant causes the time scale of the exponentially fast reduction process  to decrease proportional to $1/N$ (see the scaling rule (\ref{scaling}), and  remarks in the end of sect.\ref{bandyn}). This suggests a possibility of  retrieving the ideal measurement of position by a suitable limit process.
  \par\vskip0.2cm\noindent

  \section{Concluding Remarks.}

Dynamical instability in  Smilansky's model is due to a positive feedback loop between fall of the particle in the $\delta$- potential well and excitation of the oscillator. This effect may not crucially rest on point interaction, nor  on linear dependence of the interaction
on the coordinate of the oscillator. While such special features are probably optimal in simplifying  mathematical analysis, a similar behavior may be reproducible with  smoother interaction potentials and also  in  purely classical models.\\
Smilansky's model is somewhat unrealistic from a physical viewpoint, as it is not easy to conceive of physical
realizations, albeit approximate. Generalizations of the model  to higher dimension, and more realistic
couplings - if at all possible - may enhance physical interest.
\par\vskip 1cm\noindent
{\bf Acknowledgment:
  } I thank Uzy Smilansky for discussions about his model  and Raffaele Carlone for making me aware of exact results in related fields.

\section{Notes, and proofs.}

\subsection{BA\&WL Theory.}
\label{B&A}

  Corollary \ref{mostimp} to Proposition \ref{hold}, which is proven in this Note, is an essential ingredient in the derivation  of the spectral expansion in Sect. \ref{specexp} (notably in the proof of Lemma \ref{normeig}). Proposition \ref{hold} is proven by a rephrased version of a method which was introduced by Wong and Li \cite{WL92} in the context of a theory of Birkhoff and Adams about asymptotic expansions for $n\to+\infty$ of solutions of 2nd order difference equations of the form :
\begin{equation}
\label{n0}
C(n+2)\;+\;p(n)\;C(n+1)\;+\;q(n)\;C(n)\;=\;0\;.
\end{equation}
 The main result of that theory (Theorem 8.36 in ref.\cite{El99}) is that , whenever  coefficients $p(n)$ and $q(n)$ have asymptotic expansions for $n\to\infty$ in powers of $n^{-1}$:
\begin{equation}
\label{n1}
p(n)\;\sim\;\sum\limits_{k=0}^{+\infty}a(k)\;n^{-k}\;,\;\;\;q(n)\;\sim\;\sum\limits_{k=0}^{+\infty}b(k)\;n^{-k}\;,
\end{equation}
the equation has two linearly independent "normal" solutions, which have asymptotic expansions:
\begin{equation}
\label{n2}
C_{\pm}^{\star}(n)\;\sim\;\sigma_{\pm}^n\;n^{\alpha_{\pm}}\;\sum\limits_{s=0}^{\infty}c_{\pm}(s)\;n^{-s}
\end{equation}
where $\sigma_{\pm}$ are the (assumedly distinct) roots of the equation:
\begin{equation}
\label{n3}
\sigma^2\;+\;a(0)\;\sigma\;+\;b(0)\;=\;0\;,
\end{equation}
and
\begin{equation}
\label{n4}
\alpha_{\pm}\;=\;\frac{a(1)\;\sigma_{\pm}\;+\;b(1)}{a(0)\;\sigma_{\pm}\;+\;2b(0)}\;.
\end{equation}
Coefficients $c_{\pm}(s)$ in (\ref{n2}) are recursively determined by directly substituting (\ref{n2}) in (\ref{n0}) with $c_{\pm}(0)=1$. Eqn.(\ref{bkad1}) may be written in the form of eqn. (\ref{n0}), with coefficients that additionally depend on $E$:
\begin{equation}
\label{n11}
p(n)\;=\;p(n,E)\;=\;-\;\frac{h_{1}(n,E)}{h_2(n,E)}\;,\;\;\;q(n)\;=\;q(n,E)\;=\;-\;\frac{h_0(n,E)}{h_2(n,E)}\;,
\end{equation}
where $h_0,\;h_1,\;h_2$ are as in eqs. (\ref{p0p1}) and (\ref{p1p2}). Using (\ref{geneigf1}) and (\ref{geneigf2}) one computes  asymptotic expansions (\ref{n1}). In particular,
$$
a(0,E)\;=\;\tfrac{2\omega}{\alpha}\;,\;\;a(1,E)\;=\;-\tfrac{\omega}{\alpha}\bigl(1+\tfrac E{\omega}\bigr)\;,\;\;
b(0,E)\;=\;1\;,\;\;b(1,E)\;=\;-1\;,
$$
 whence it follows that expansions (\ref{n2}) have the form (\ref{bkad2}) at lowest orders.\\
 In the following  $I$ will denote an arbitrary closed interval contained in $\RM\setminus{\cal E}^{*}_{\omega}$;   positive quantities only dependent on $\alpha,\omega,I$ will be denoted by $c_1,c_2,\ldots$; derivatives with respect to $E$ will be denoted by a dot, like, {\it e.g.},  in $\dot{p}(n,E),\dot{q}(n,E)\ldots$. The following Lemma \ref{uniasy} sets premises for the proof of  Proposition  \ref{hold}.
\begin{lemma}
\label{uniasy}
{For all positive $n$, $p(n,E)$ and $q(n,E)$ as given by (\ref{n11}), (\ref{p0p1}), (\ref{p1p2})  are $C^1$ functions of $E\in I$. Their  asymptotic expansions (\ref{n1}) are uniform in $I$ . Their derivatives have uniform asymptotic expansions in $I$ in powers of $n^{-1}$, with coefficients given by the derivatives of the  coefficients $a(s,E)$, $b(s,E)$, as specified in (\ref{n11}), (\ref{p0p1}), and (\ref{p1p2}). In particular,  $n|\dot{p}(n,E)|$ and $n|\dot{q}(n,E)|$ are bounded in $I$ by some $c_1>0$.
}
\end{lemma}
{\it Proof}: by direct inspection. $\Box$\\
\begin{proposi}
\label{hold}
{ For all $n\geq 0$ the normal solutions $C_{\pm}^{\star}(n,E)$ of eqn.(\ref{n0}), with coefficients as in (\ref{n11}), are $C^1$  functions of $E\in I$.  }
\end{proposi}
{\it Proof}: the proof  is the same for both normal solutions, so suffixes $\pm$ will be left understood throughout. Thanks to Lemma \ref{uniasy}, all coefficients
$c(s,E)$ are $C^1$ functions of $E\in I$, because each of them is determined by a finite number of coefficients $a$ and $b$.
Let $\mathfrak R$ and ${\mathfrak R}_0$ denote operators that act on sequences $w: \NM\times\RM\to\CM$ according to
\begin{gather}
({\mathfrak R}w)(n,E)\;=\;w(n+2,E)\;+\;p(n,E)w(n+1,E)\;+\;q(n,E)w(n,E)\;,\\
({\mathfrak R}_0w)(n,E)\;=\;w(n+2,E)\;+\;a(0,E)w(n+1,E)\;+\;b(0,E)w(n,E)\;.
\end{gather}
Let $Y(n,E)\;=\;\exp(\pm in\theta\mp i\lambda E\log(n))$, and let a normal solution be written in the form:
\begin{equation}
\label{n5}
C^{\star}(n,E)=L_{N}(n,E)+\epsilon_N(n,E)\;,
\end{equation}
where $N$ is an  integer, and $L_N(n,E)$ is obtained on truncating at the $(N-1)$-th order the asymptotic expansion (\ref{n2}) of the normal solution:
$$
L_{N}(n,E)=n^{-1/2}Y_{}(n,E)\sum_0^{N-1}c_{}(s,E)n^{-s}\;.
$$
Direct calculation yields
\begin{equation}
\label{n6}
({\mathfrak R}L_{N})(n,E)\;=\;n^{-1/2}\;Y(n,E)\;R_{N}(n,E)\;,
\end{equation}
where , for any fixed $n$, $R_{N}(n,E)$ is a $C^1$ function of $E\in I$, because such are all coefficients $c(s,E)$, $(0\leq s\leq N-1)$; and, moreover,
\begin{equation}
\label{n61}
R_{N}(n,E)=O(n^{-N-1})\;,
\end{equation}
uniformly with respect to $E\in I$ as $n\to\infty$. Differentiating (\ref{n6}) on both sides,  $\dot{R}_N(n,E)$ is found to have a uniform  asymptotic expansion in $I$ in powers of $n^{-1}$, so (\ref{n61}) entails that
\begin{equation}
\label{n62}
\dot{R}_{N}(n,E)=O(n^{-N-1})\;,
\end{equation}
uniformly in $I$. Substitution of  (\ref{n5}) and (\ref{n6}) into (\ref{n0}) yields:
$$
({\mathfrak R}\epsilon_N)(n,E)\;=\;-n^{-1/2}\;Y(n,E)\;R_N(n,E)\;,
$$
which is equivalent to
\begin{equation}
\label{n66}
({\mathfrak R}_0\epsilon_N)(n,E)\;=\;-n^{-1/2}\;Y(n,E)\;R_N(n,E)\;-\;{\tilde q}(n,E)\epsilon_N(n,E)\;-\;
{\tilde p}(n,E)\epsilon_N(n+1,E)\;,
\end{equation}
where ${\tilde p}(n,E)=p(n,E)-a_0(E)$ and ${\tilde q}(n,E)=q(n,E)-b_0(E)$\;. Eqn.(\ref{n66}) may be read as a inhomogeneous 2nd order difference equation, so it can be rewritten in "integral" form using a "Green function" for the operator ${\mathfrak R}_0$. This is provided by the function \cite{WL92}:
\begin{equation}
\label{n7}
G(n)\;=\;s(n-1)\;\frac{\sin\bigl((n-1)\theta\bigr)}{\sin(\theta)}
\end{equation}
where $s(n)=1$ for $n\geq 0$, and $s(n)=0$ for $n<0$. Therefore, introducing the operator $\mathfrak G$ that formally acts on sequences as in:
\begin{equation}
\label{defopk}
({\mathfrak G}w)(n,E)\;=\;-\sum\limits_{k=n}^{+\infty}G(n-k)\;\bigl({\tilde q}(k,E)\;w(k,E)\;+\;{\tilde p}(k,E)\;w(k+1,E)\bigr)\;,
\end{equation}
and denoting
\begin{equation}
\label{dienne}
d_N(n,E)\;=\;-\sum\limits_{k=n}^{+\infty}G(n-k)\;k^{-1/2}Y(k,E)R_N(k,E)\;,
\end{equation}
the sequence $\epsilon_N(n,E)$ must solve the following equation written in vector form :
\begin{equation}
\label{n8}
\epsilon_N\;=\;d_N\;+\;{\mathfrak G}\;\epsilon_N\;.
\end{equation}
No solution of the homogeneous equation ${\mathfrak R}_0\epsilon_N=0$  appears on the rhs of (\ref{n8}), because such solutions do not vanish at infinity, as is instead required of $\epsilon_N(n,E)$. Let $\sigma$ and $\sigma^{\dagger}$  denote the left shift operator
and its adjoint: $(\sigma w)(n,E)=w(n+1,E)$, $(n\geq 1)$, $(\sigma^{\dagger}w)(n,E)=w(n-1,E)$ if $n>1$, and $(\sigma^{\dagger}w)(1,E)=0$.
If  $\epsilon_N$ satisfies (\ref{n8}), then ${\tilde\epsilon}_{N}:=\sigma^{N}\epsilon_N$ satisfies:
\begin{equation}
\label{n9}
{\tilde\epsilon}_{N}\;=\;\sigma^{N}d_N\;+\;{\mathfrak G}_{N}\;{\tilde\epsilon}_{N}\;,\;\;\;\;{\mathfrak G}_N\;=\;\sigma^N{\mathfrak G}\sigma^{\dagger N}\;.
\end{equation}
The operator ${\mathfrak G}_N$ is explicitly given by eqn.(\ref{defopk}) after replacing $\tilde a$, $\tilde b$ by
 $\sigma^N{\tilde a}$, $\sigma^N{\tilde b}$ respectively.
Thanks to Lemma \ref{d1} and to the  Contraction Mapping theorem,  if $N$ is sufficiently large then
eqn.(\ref{n9}) has a unique solution in
the Banach space ${\mathfrak X}_{N,I}$ of sequences $w:\NM\to C^1(I)$ such that
\begin{equation}
\label{norms}
\|w\|_{\mathfrak X}\;:=\;\|w\|_{N}\;+\;\|\dot{w}\|_{N*}\;<\;+\infty\;,
\end{equation}
where
\begin{gather}
\|w\|_{N}\;=\;\sup\;\{(N+n)^{N+1/2}|w(n,E)|\;,\;n\geq 0,\;E\in I\}\;,\nonumber\\
\|w\|_{N^*}\;=\;\sup\;\biggl\{(N+n)^{N+1/2}\frac1{\log(N+n)}|w(n,E)|\;,\;n\geq 0,\;E\in I\biggr\}\;.
\end{gather}
The thus found ${\tilde\epsilon}_{N}$ determines $\epsilon_N(n,E)$ as a $C^1$ function, and hence, via eqn.(\ref{n5}), the normal solution, for $n\geq N+1$. For such $n$ the thesis  is then proven, because $L_N(n,E)$ is itself $C^1$ wrt $E$ thanks to already noted properties of coefficients $c(s,E)$. The values of the normal solution thus found at $n=N+1$ and $n=N+2$  can then be used to retrieve the normal solution for
$0\leq n\leq N$ by solving eqn.(\ref{n0}) backwards (which is possible, because $q(n,E)\neq 0$ for all $n\geq 0$, thanks to the assumption that $E$ is not in ${\cal E}_{\omega}^*$). As this process involves a finite number of steps, and $p,q$ are $C^1$ functions, the proof is  complete. $\Box$\\
\begin{lemma}
\label{d1}
{(i)  $\sigma^{N}d_N\in{\mathfrak X}_{N,I}$, (ii) ${\mathfrak G}_N$ is a bounded operator in ${\mathfrak X}_{N,I}$,
and its norm is bounded by:
$$
\|{\mathfrak G}\|_{\mathfrak X}\;\leq\;2ec_2(c_1+3\beta)(N+1/2)^{-1}\;,
$$
where $c_1$ is as in Lemma \ref{uniasy}, $c_2=\sin(\theta)^{-1}=(1-\omega^2/\alpha^2)^{-1/2}$, and
$$
\beta\;=\;\sup\;\{k\;\bigl(|{\tilde p}(k,E)|\;+\;|{\tilde q}(k,E)|\;,\;k\in\NM,\;E\in I\bigr\}\;.
$$
}
\end{lemma}
{\it Proof}: (i) from eqs.(\ref{dienne}) and  (\ref{n61}):
$$
d_N(n,E) \;\leq\;c_2\sum\limits_{k=n}^{+\infty}k^{-N-3/2}\;=\;O(n^{-N-1/2})\;,
$$
therefore $\|d_N\|_{N}$ is finite and so is $\|\sigma^{N+1}d_N\|_{N}$. Next, the derivative of the $k$-th term in the sum on the rhs in eqn.(\ref{dienne}) is $O(k^{-N-3/2}\log(1+k))$, so the sum of such derivatives is absolutely and uniformly convergent in $I$ to the derivative of $d_N(n,E)$, and $\|\dot{d}_{N}\|_{N*}$ is finite.\\
(ii): noting that
$$
\sup\{(k+N)\sigma^N\tilde{p}(k,E)\;|\;k\ge n\}\;\leq\;\beta\;,
$$
and similarly for $\tilde q$, one may write:
\begin{gather}
\bigl|({\mathfrak G} _N w))(n,E)\bigr|\;\leq\;c_2\beta\|w\|_{N}\sum\limits_{k=n}^{+\infty}(N+k)^{-N-3/2}\nonumber\\
\leq\;c_2\beta\|w\|_N\int_{n-1}^{+\infty}\;(N+x)^{-N-3/2}\nonumber\\
=\;c_2\beta\|w\|_N(N+1/2)^{-1}(N+n-1)^{-N-1/2}\;,
\label{norm1}
\end{gather}
so
\begin{equation}
\label{half}
\|{\mathfrak G}_Nw\|_N\;\leq \;2c_2e\beta(N+1/2)^{-1}\|w\|_N
\end{equation}
 thanks to $(N+n)^{N+1/2}(N+n-1)^{-N-1/2}<2e$.
To estimate $\|({\mathfrak G} _N w)^{\dot{}}\|_{N*}$:
\begin{gather}
({\mathfrak G} _N w)^{\dot{}}(n,E)\;\leq\;c_2c_1\|w\|_N\sum\limits_{k=n}^{+\infty}(k+N)^{-N-3/2}\;+\;\nonumber\\
+\;c_2\beta\|\dot{w}\|_{N*}\sum\limits_{k=n}^{+\infty}(k+N)^{-N-3/2}{\log(k+N)}\;.
\end{gather}
Estimating the sums on the rhs as it was done in (\ref{norm1}) leads to:
$$
\|({\mathfrak G} _N w)^{\dot{}}\|_{N*}\;\leq\;\bigl(2c_1c_2e\|w\|_N\;+\;4c_2e\beta\|\dot{w}\|_{N*}\bigr)(N+1/2)^{-1}\;.
$$
Thanks to definition (\ref{norms}), the latter estimate along with  (\ref{half}) yield the claimed bound on the norm of ${\mathfrak G}_N$ as an operator
in ${\mathfrak X}_{N,I}$. $\Box$\\
\begin{coro}
\label{mostimp}
{If $E\in\RM\setminus{\cal E}_{\omega}^*$ then the difference equation  (\ref{bkad1}) has a particular solution which  satisfies the initial condition (\ref{condin}), and moreover has the asymptotics (\ref{asycn}), where $\zeta(E)$ is a $C^1$ function of $E$ in any closed interval  of energies containing no exceptional points.}
\end{coro}
{\it Proof:}
the complex amplitude $Z(E)$ (cp. eqn.(\ref{nmfct})), which determines the sought for solution in terms of the normal solutions, is found by solving eqs.(\ref{zeta}); so it is a smooth function  of the values of the normal solutions at $n=0$ and $n=1$. The conclusion follows, because $\zeta(E)$ is the phase of $Z(E)$.  $\Box$

\subsection{Proof of Lemma \ref{specmap}.}
\label{proofspecmap}
First it will be proven that if $E^m\Psi(E)\in L^2(\RM)$ for some integer $m$ then $\psi=\imath(\Psi)$ is in the domain of ${\cal H}^m$, and $\imath(E^m\Psi)={\cal H}^m{\psi }$.\\
It is easy to see that
$$
L_n^m\psi_n(x)=c\int_{\RM}dE\;\Psi(E)L^m_nu_n(x,E)=c\int_{\RM}dE\;E^m\Psi(E)u_n(x,E)
$$
(with $L_n$ defined as in  (\ref{ellenne}))  holds for all $\Psi\in C_0(\RM\setminus{\cal E}_{\omega}^*)$ and
all positive integers $m,n$; so, thanks to
(\ref{orth})
the sequence $\{L_n^m\psi_n\}_n$ is in $\tp$ whenever $E^m\Psi(E)\in L^2(\RM)$. On the other hand the sequence $\{L^m_nu_n\}_{n}$ satisfies the matching condition (\ref{flux}) because so does $u_n$, and because  $L_nu_n=Eu_n$. The same is  then true of the  sequence $\{L^m_n\psi_n\}_n$, because the condition $\Psi\in C_0(\RM\setminus{\cal E}_{\omega}^*)$ allows for computing left- and right-hand derivatives of $L^m\psi_n(x)$ at $x=0$ under the integral sign . Therefore $\psi$ is in the domain of ${\cal H}^m$ whenever $\Psi\in C_0(\RM\setminus{\cal E}_{\omega}^*)$, and ${\cal H}^m\psi=\imath(E^m\Psi)$. As  ${\cal H}^m$ is a closed operator, the same is true whenever $\Psi\in L^2(\RM)$ and $E^m\Psi\in L^2(\RM)$.  \\
(ii) follows by continuity, because $\imath$ is isometric. \\
(iii) To prove that $\imath$ is onto: thanks to (\ref{orth}) and (\ref{specth}), the time-correlation $(\psi,e^{-i{\cal H}t}\psi)$ coincides with the Fourier transform of $|\Psi(E)|^2$. Therefore, $|\Psi(E)|^2$ is
the density of the absolutely continuous spectral measure of $\psi$ with respect to ${\cal H}$ (also known as the local density of states). As ${\cal H}$ has  a simple absolutely continuous spectrum coinciding with $\RM$, $\psi$ is a cyclic vector whenever its local density of states is Lebesgue-almost everywhere different from zero. So, whenever $\Psi(E)$ is a.e. nonzero, $\psi$ is a cyclic vector of ${\cal H}$, so the closed span of $\{e^{-i{\cal H}t}{\psi}\}_{t\in\RM}$  is the whole of $\tp$, whence   $\tp=\imath(L^2(\RM))$ follows.\\

\subsection{ Proof of Proposition \ref{expeng}.}
\label{app1}
The expectation value of the energy of the oscillator in a state $\psi=\{\psi_n(x)\}\in\tp$ of the composite system is given by:
$$
\bigl(\psi, \mathbb{I}\otimes\hosc\;\psi\bigr)\;=\;\sum\limits_{n=0}^{\infty}(n+\tfrac12)\omega\int_{-\pi}^{\pi}dx\;
|\psi_n(x)|^2\;,
$$
and so the sequence
\begin{equation}
\label{defpn}
P(n,E)\;:=\;\int_{-\pi}^{\pi}dx\;|u_n(x,E)|^2\;,\;\;\;(n\geq 0)\;,
\end{equation}
may be thought of as a non-normalizable distribution  of the energy of the oscillator over its unperturbed levels, when the full system has energy $E$. The present proof of Proposition \ref{expeng} rests on the following inequality, which is an immediate consequence of eqs. (\ref{defC}),(\ref{geneigf1}), (\ref{geneigf2}), and (\ref{asycn}):
\begin{equation}
\label{estpn}
\limsup\limits_{n\to+\infty}\;nP(n,E)\;\leq\;\pi^{-1}\;.
\end{equation}
Let $\psi\in{\mathfrak H}_{\mbox{\rm\tiny ac}}$,  $\|\psi\|=1$, and $\Psi(E)$ its spectral representative, so that $|\Psi(E)|^2$ is the
density of the absolutely continuous spectral measure of $\psi$ with respect to ${\cal H}$. Thanks to (\ref{estpn})
there is a continuous function $\Psi_1(E)$ , compactly supported in $\RM\setminus{\cal E}_{\omega}^* $,  so that on the one hand:
\begin{equation}
\label{18}
\int dE\;\bigl|\Psi(E)-\Psi_1(E)\bigr|^2<1/8\;,
\end{equation}
and on the other hand  :
\begin{equation}
\label{allpsi}
\sum_{n=0}^N P(n,E)<C\ln(N)\;,\;\;\; \forall N\in\NM\;
\end{equation}
for some positive constant $C$ and for all $E$ in the support of $\Psi_1$. Let $\psi_1=\imath(\Psi_1)$,  $\psi_2=\psi-\psi_1$, and  $\psi(t)=e^{-i{\cal H}t}\psi
=\{\psi_n(x,t)\}\in\tp$. For $T>0$ , the probability of finding the energy of the oscillator in its $n$-th level, averaged from time $0$ to time $T$, is:
\begin{gather}
\label{timeavpr}
p_n(T)=\frac1T\int_0^Tdt\;\int_{-\pi}^{\pi}dx\;|\psi_n(x,t)|^2\;.
\end{gather}
Let $p_{1,n}(T)$ and $p_{2,n}(T)$ denote the functions which are defined by the same equation,
with $\psi$ replaced by $\psi_1$ and $\psi_2$ respectively. By construction of $\Psi_1$, the spectral representation
of $\psi_1$ has the form (\ref{cint}), so Proposition \ref{specmap} yields:
\begin{equation}
\label{kl1}
p_{1,n}(T)\;=\;\frac1T\int_0^Tdt\;\int_{-\pi}^{\pi}dx\;\biggl|\int dE\;e^{-iEt}\Psi_1(E)\;u_n(x,E)\biggr|^2\;,
\end{equation}
On the other hand, denoting $\cal{F}[.]$ the Fourier-Plancherel transform in $L^2(\RM)$:
\begin{gather}
\int_0^Tdt\;\biggl|\int dE\;e^{-iEt}\Psi_1(E)\;u_n(x,E)\biggr|^2\;=\;2\pi\int_0^Tdt\;
\bigl|{\cal F}\bigl[\Psi_1\;u_n(x,.)\bigr](t)\bigr|^2\\
\leq\;2\pi\|{\cal F}\bigl[\Psi_1\;u_n(x,.)\bigr]\|^2\;=\;2\pi\|\Psi_1\;u_n(x,.)\|^2\\
=\;2\pi\int dE\;u_n^2(x,E)|\Psi_1(E)|^2\;.
\label{kl2}
\end{gather}
Replacing (\ref{kl2}) in (\ref{kl1}), and using (\ref{defpn}) and inequality (\ref{allpsi}), which holds throughout the support of $\Psi_1$ :
\begin{equation}
\label{kl3}
\sum\limits_{n=0}^{N}p_{1,n}(T)\;
\leq\;2\pi CT^{-1}\ln(N)\;.
\end{equation}
With $N=N(T):=e^{C_1T}$, where $C_1=1/(16\pi C)$, this estimate yields:
\begin{equation}
\label{kl4}
\sum\limits_{n\leq N_T}p_{1,n}(T)\;<\;\tfrac18.
\end{equation}
From the definitions of $p_n(T),p_{1,n}(T),p_{2,n}(T)$, and ineq. (18)  it immediately follows that:
$$
\sum\limits_{n=0}^{N(T)}p_n(T)\;\leq\;2\sum\limits_{n=0}^{N(T)}p_{1,n}(T)\;+\;
2\sum\limits_{n=0}^{N(T)}p_{2,n}(T)\;<\;
\tfrac14\;+\;\tfrac14\;=\;\tfrac12\;,
$$
and so
\begin{equation}
\label{almostfin}
\frac1T\int_0^Tdt\;\bigl(\psi(t), \mathbb{I}\otimes\hosc\;\psi(t)\bigr)\;>\;
\bigl(\tfrac12\;+\;N(T)\bigr)\omega\sum\limits_{n>N(T)}p_n(T)\;>
\;\tfrac12N(T)\omega\;=\;\tfrac12e^{C_1T}\omega\;.
\end{equation}
\subsection{ Proof of Proposition \ref{collaps}.}
\label{app2}
For $0<\eta<\pi$ let $A_{\eta}=(-\pi,-\eta)\cup(\eta,\pi)$. The explicit form of $u_n$ given in eqs.(\ref{defC}) and (\ref{geneigf1}) shows that
$|u_n(x)|\leq|C(n,E)|\rho_n\cosh(\chi_n(E)(\pi-\eta))$, whenever $n>E/\omega-1/2$ and $x\in A_{\eta}$; so, if in addition  $E\in\RM\setminus\exc$, then from (\ref{geneigf2}) and from  the asymptotic formula (\ref{asycn}) it follows  that
\begin{equation}
\label{cvae}
\sum\limits_{0}^{+\infty}\int_{A_{\eta}}dx\;u_n^2(x,E)\;<\;+\infty\;.
\end{equation}
because the integrals in the sum decrease exponentially
fast for $n\to\infty$.  Thanks to (\ref{cvae}), for  $0<\epsilon<1$ one can find a compact set $B_{\epsilon}\subset\RM\setminus\exc$ and a continuous function $\Psi_{\epsilon}$ supported in $B_{\epsilon}$, so that, on the one hand:
\begin{equation}
\label{bdsq}
\sum\limits_{0}^{+\infty}\int_{A_{\eta}}dx\;u_n^2(x,E)\;<\;C_{\epsilon}\;,\;\;\;\forall E\in B_{\epsilon}\;,
\end{equation}
for some positive constant $C_{\epsilon}$; and, on the other hand, $\|\psi-\psi^{(\epsilon)}\|^2<\epsilon$, where $\psi^{(\epsilon)}=\imath(\Psi_{\epsilon})$.  Then:
\begin{gather}
\iint\limits_{A_{\eta}\times\RM}dx\;dq\;|\psi(x,q,t)|^2\;\leq\;2\iint\limits_{A_{\eta}\times\RM}
\;|\psi(x,q,t)-\psi^{(\epsilon)}(x,q,t)|^2\;+\nonumber\\
+\;2\iint\limits_{A_{\eta}\times\RM}dx\;dq\;|\psi^{(\epsilon)}(x,q,t)|^2\nonumber\\
\leq\;2\|\psi-\psi^{(\epsilon)}\|^2\;+\;2\iint\limits_{A_{\eta}\times\RM}dx\;dq\;|\psi^{(\epsilon)}(x,q,t)|^2\nonumber\\
\leq\;2\epsilon\;+\;2\iint\limits_{A_{\eta}\times\RM}dx\;dq\;|\psi^{(\epsilon)}(x,q,t)|^2\;.
\label{psi1psi2}
\end{gather}
From eqn.(\ref{herexp}):
$$
\iint\limits_{A_{\eta}\times\RM}dx\;dq\;|\psi^{(\epsilon)}(x,q,t)|^2\;=\;\sum\limits_{n=0}^{+\infty}\int_{A_{\eta}}dx\;
|\psi^{(\epsilon)}_n(x,t)|^2\;,
$$
and then, since the spectral representative $\Psi_{\epsilon}$ of $\psi^{(\epsilon)}$ is compactly supported in $\RM\setminus\exc$, the spectral representation (\ref{cint}) can be used to the effect that:
\begin{gather}
\label{cvae1}
\iint\limits_{A_{\eta}\times\RM}dx\;dq\;|\psi^{(\epsilon)}(x,q,t)|^2\;=\;
\iint\limits_{B_{\epsilon}\times B_{\epsilon}} dE\; dE'\;e^{-i(E'-E)t}\;\;\overline{\Psi_{\epsilon}(E)}\Psi_{\epsilon}(E')\;G_{\eta}(E,E')\;,
\end{gather}
where :
$$
G_{\eta}(E,E')\;=\;\int_{A_{\eta}}dx\;\sum\limits_{n=0}^{+\infty}\;u_n(x,E)u_n(x,E')\;.
$$
Thanks to (\ref{bdsq}), $G_{\eta}(E,E')$ is bounded in $B_{\epsilon}\times B_{\epsilon}$; on the other hand
$\Psi_{\epsilon}(E)$ is summable over $B_{\epsilon}$, so the integral on the rhs in
(\ref{cvae1}) tends to $0$ in the limit $t\to\infty$ thanks to the Riemann-Lebesgue lemma. From
 (\ref{psi1psi2}) it follows that
$$
\limsup\limits_{t\to\infty}\iint\limits_{A_{\eta}\times\RM}dx\;dq\;|\psi(x,q,t)|^2\;\leq\;2\epsilon\;,
$$
whence the claim (\ref{delta}) follows, because $\epsilon>0$ is arbitrary.

\subsection{Proof of Corollary \ref{dec}.}
\label{proofcor}
By positivity of $S(t)$, it is sufficient  to prove the claim for $\phi=\phi'$ and $\|\phi\|=1$. Let $P_{\phi}=
(\phi,.)\phi$ denote projection along $\phi$, and let $P_{\eta}$, $P^{\bot}_{\eta}$  respectively denote projection onto the functions supported in $\eta<|x|<\pi$, and its orthogonal complement. Then:
\begin{equation}
\label{trto0}
(\phi, S(t)\phi)\;=\;\Tr(S(t)P_{\phi})\;\leq\;|\Tr(S(t)P_{\phi}P_{\eta})|\;+\;|\Tr(S(t)P_{\phi}P^{\bot}_{\eta})|\;.
\end{equation}
For any $\eta>0$ the 1st term on the rhs in (\ref{trto0}) tends to $0$ as $t\to\pm\infty$ thanks to eqn.(\ref{delta}), due  to
$$
|\Tr(S(t)P_{\phi}P_{\eta})|\;=\;|\bigl(\psi(t),P_{\phi}P_{\eta}\otimes{\mathbb I}\; \psi(t)\bigr)|\;\leq\;\|P_{\eta}\otimes{\mathbb I}\;\psi(t)\|\;.
$$
On the other hand, the 2nd term on the rhs in (\ref{trto0}) can be made arbitrarily small, uniformly with respect to $t$,  by choosing $\eta$ small enough:
$$
|\Tr(S(t)P_{\phi}P^{\bot}_{\eta})|\;\leq\;\|P_{\phi}P^{\bot}_{\eta}\|\;\leq\;\|P_{\eta}^{\bot}\phi\|\;,
$$
 Hence the lhs in (\ref{trto0}), which does not depend on $\eta$,  tends to $0$ in the limit $t\to\pm\infty$.

\subsection{Proof of Proposition \ref{BO}.}
\label{BOproof}

\begin{gather}
\int_{\RM}dq\;|Q_n(q,t)|^2\;=\;\int_{\RM}dq\;\biggl|
\biggl(\int_{|x|<\eta}+\int_{\eta<|x|<\pi}\biggr) dx\;\phi_{q,n}(x)\psi(x,q,t)\biggr|^2\nonumber\\
\leq\;\int_{\RM}dq\;\bigl\{R(\eta,q)\;+\;S(\eta,q)\bigr\}
\end{gather}
where:
\begin{eqnarray}
R(\eta,q)\;&=&\;2\;\biggl|\int_{|x|<\eta}dx\;\phi_{q,n}(x)\psi(x,q,t)\biggr|^2\nonumber\\
S(\eta,q)\;&=&\;2\;\biggl|\int_{\eta<|x|<\pi}dx\;\phi_{q,n}(x)\psi(x,q,t)\biggr|^2
\end{eqnarray}
From the Cauchy-Schwarz inequality:
\begin{gather}
\int_{\RM}dq\;R(\eta,q)\;\leq\;2\int_{\RM}dq\biggl(\int_{|x|<\eta}dx\;\phi^2_{q,n}(x)\biggr)\biggl(
\int_{|x|<\eta}dx\;|\psi(x,q,t)|^2\biggr)\;,
\end{gather}
and using that $\phi_{q,n}$ with $n>0$ are uniformly bounded (by $(\pi-1)^{-1/2}$) and that $\|\psi(t)\|=1$,
$$
\int_{\RM}dq\;R(\eta,q)\;\leq\;C\eta\;,
$$
for a suitable constant $C$. Similarly, using Cauchy-Schwarz and $\int dx\phi_{q,n}^2(x)=1$,
$$
\int_{\RM}dq\;S(\eta,q)\;\leq\;\int_{\RM}dq\int_{\eta<|x|<\pi}dx\;|\psi(x,q,t)|^2\;,
$$
so, thanks to Proposition \ref{collaps},
$$
\limsup\limits_{t\to\infty}\int_{\RM}dq\;|Q_n(q,t)|^2\;\leq\;C\eta\;,
$$
and the claim follows because $\eta>0$ is arbitrary. $\Box$\\
The above argument fails if $n=0$, because $\phi_{q,0}$ is not uniformly bounded in $q<0$.

\subsection{About the Band potential.}
\label{boextra}

Here the 3d term in the band potential (\ref{bandpt}) is estimated. A standard perturbative calculation yields:
\begin{equation}
\label{adia1}
\gamma_{nl}(q)\;\equiv\;\int_{\SM}dx\;\phi_{q,n}(x)\frac{\partial \phi_{q,l}(x)}{\partial q}\;=\;
\alpha\;\frac{\phi_{q,n}(0)\phi_{q,l}(0)}{W_l(q)-W_n(q)}\;,\;\;\;(n\neq l)
\end{equation}
and $\gamma_{nn}=0$; so, thanks to orthonormality and completeness of $\{\phi_{q,n}\}$:
\begin{gather}
\label{asybo}
\int_{\SM}dx\;\biggl(\frac{\partial\phi_{q,0}}{\partial q}\biggr)^2\;=\;\sum\limits_{n=1}^{+\infty}
\gamma_{n0}^2(q)\\
=\;\alpha^2\phi_{q,0}^2(0)\sum\limits_{n=1}^{+\infty}\frac{\phi_{q,n}^2(0)}{(W_n(q)-W_0(q))^2}\;.
\end{gather}
It is easy to see that $|\phi_{q,n}(0)|\leq(\pi-1)^{-1/2}$ whenever $n>0$, that $W_n(q)>\tfrac12n^2$, and that
$|\phi_{q,0}|\sim\sqrt{-\alpha q}$ for $q\to-\infty$. Using this and the asymptotic
form of $W_0(q)$ given in (\ref{gren}), (\ref{asybo}) is found to be $O(q^{-2})$ for $q\to-\infty$ and $\sim$const. as $q\to+\infty$.
\subsection{Proof of Proposition \ref{multicoll}.}
\label{proofmulticoll}
The following notations will be used. For sufficiently small $\eta>0$,
 $A_{j,\eta}=J_j\setminus D_j$ where
$D_j\subset J_j$ is an arc of size $\eta>0$ centered at $O_j$; and  $A_{\eta}=\cup_jA_{j,\eta}$.  $P_{\eta}$ will denote the
projector of $L^2(\SM)\otimes L^2(\RM^N)$ onto $L^2(A_{\eta})\otimes L^2(\RM^N)$, and $P_{j,\eta}$ will denote the projector of $L^2(J_j)\otimes L^2(\RM^N)$ onto $L^2(A_{j,\eta})\otimes L^2(\RM^N)$. \\
Existence of $\Omega_{\pm}\equiv \Omega_{\pm}({\cal H}^{(b)},{\cal H})$
entails that:
\begin{gather}
\lim\limits_{t\to\pm\infty}(e^{-i{\cal H}t}\psi\;, P_{\eta}\;e^{-i{\cal H}t}\psi)\;=\;
\lim\limits_{t\to\pm\infty}(e^{-i{\cal H}^{(b)}t}\Omega_{\pm}\psi\;, P_{\eta}\;e^{-i{\cal H}^{(b)}t}\Omega_{\pm}\psi)\nonumber\\
=\;\lim\limits_{t\to\pm\infty}\sum\limits_{j=1}^N(e^{-i{\cal H}^{(b)}t}\Omega_{\pm}\psi\;, P_{j,\eta}\;e^{-i{\cal H}^{(b)}t}\Omega_{\pm}\psi)\;.
\end{gather}
The quantity of which the $t\to\infty$ limit is taken in the above equations  is the probability of finding the particle in $A_{\eta}$ at time $t$.
Each subspace $P_j({\mathfrak H})$ is invariant under the evolution ruled by ${\cal H}^{(b)}$, so the sum in the last line is equal to
\begin{gather}
\sum\limits_{j=1}^N(e^{-i{\cal H}^{(b)}t}P_j\Omega_{\pm}\psi\;, P_{j,\eta}\;e^{-i{\cal H}^{(b)}t}P_j\Omega_{\pm}\psi)\;.
\end{gather}
Each term in the sum is a probability of finding the particle at time $t$ in $J_{j,\eta}$ with the evolution
$e^{-i{\cal H}^{(b)}t}$. In each invariant subspace this  evolution is that of a single-oscillator model, so, thanks to Proposition \ref{collaps},
each term in the sum tends to $0$ as $t\to\infty$. Hence, so does the probability of finding the particle in $A_{\eta}$ for all $\eta>0$, which is equivalent to the thesis. $\Box$


\begin{thebibliography}{99}

\bibitem{smi04}{U.Smilansky, \textit{ Irreversible Quantum Graphs}, Waves in Random Media {\bf 14} (2004) 143.}
\bibitem{MS04}{M.Solomyak, \textit{On a differential operator appearing in the theory of irreversible quantum graphs}, Waves in Random Media {\bf 14} (2004) 173.}
\bibitem{NS06}{S.N.Naboko, M.Solomyak, \textit{ On the absolutely continuous spectrum in a model of an irreversible quantum graph}, Proc. London Math. Soc. (3) {\bf 92} (2006) 251.}
\bibitem{ES05}{W.D.Evans, M.Solomyak, \textit{ Smilansky's model of irreversible quantum graphs: I. The absolutely continuous spectrum} J.Phys A (Math. Gen.) {\bf 38} (2005) 4611.}
\bibitem{CCF07}{C.Cacciapuoti, R.Carlone, and R.Figari, \textit{ A solvable model of a Tracking Chamber}, Rep. Math. Phys.
{\bf 59} (2007) 337.}
\bibitem{CCF05}{C.Cacciapuoti, R. Carlone, and R.Figari, \textit{Decoherence induced by scattering: a three-dimensional model}, J. Phys. A (Math. Gen.) {\bf 38} (2005) 4933.}
 \bibitem{DAFT08}
 {G.Dell'Antonio, R.Figari, and A.Teta, \textit{Joint excitation probability for two harmonic oscillators in one dimension and the Mott problem}, J. Math. Phys. {\bf 49} (2008) 042105.}
\bibitem{HS03}{K. Hornberger, J.E. Sipe, \textit{Collisional Decoherence reexamined}, Phys. Rev. A {\bf 68}, (2003) 012105.}
\bibitem{KL}{A. Kiselev, Y.Last, \textit{ Solutions, spectrum, and dynamics for Schr\"odinger operators on infinite domains}, Duke Math. J. Volume {\bf 102}, 1 (2000), 125-15.}
\bibitem{EMOT55}{A.Erd\'elyi, W. Magnus, F. Oberhettinger, F. Tricomi, \textit{Higher transcendental functions}. Vol. II, McGraw-Hill
(1955), p.207.}
\bibitem{RS2}{M.Reed and B.Simon, in \textit{ Methods of Modern Mathematical Physics} vol.II, Academic Press, San Diego, CA 1975, p.168 example 3.}
\bibitem{RS3}{M.Reed and B.Simon, in \textit{ Methods of Modern Mathematical Physics} vol.III, Academic Press, San Diego, CA 1975. }
\bibitem{El99}{S.N.Elaydi, \textit{An introduction to Difference Equations}, Springer New York 1999.
}
\bibitem{WL92}{R.Wong and H.Li, \textit{Asymptotic expansions for second-order difference equations}, J. Comp. Appl. Math.
{\bf 41} (1992) 65.}

\end{thebibliography}
\end{document}